\documentclass[ aip, jmp, amsmath,amssymb, reprint, ]{revtex4-1}

\usepackage{amssymb}
\usepackage{amsfonts}
\usepackage{amsmath}
\usepackage{amsthm}
\usepackage{epsfig}
\usepackage{color} 
\usepackage{subfigure}
\usepackage{enumitem}

\begin{document}

\title{Stability and control of power grids with diluted network topology}

\author{Liudmila Tumash}
\email{liudmilatumash@gmx.de}
\affiliation{Institut f{\"u}r Theoretische Physik, Technische Universit{\"a}t Berlin, Hardenbergstra\ss{}e 36, 10623 Berlin, Germany}
\author{Simona Olmi}
\email{simona.olmi@inria.fr}
\affiliation{INRIA Sophia Antipolis M\'editerran\'ee, 2004 Route des Lucioles, 06902 Valbonne, France}
\affiliation{CNR - Consiglio Nazionale delle Ricerche - Istituto dei Sistemi Complessi, 50019, Sesto Fiorentino, Italy}
\author{Eckehard Sch{\"o}ll}
\affiliation{Institut f{\"u}r Theoretische Physik, Technische Universit{\"a}t Berlin, Hardenbergstra\ss{}e 36, 10623 Berlin, Germany}

\date{\today}

\begin{abstract}
 In the present study we consider a random network of Kuramoto oscillators with inertia in order to mimic and investigate the dynamics emerging in high-voltage power grids. 
 The corresponding natural frequencies are assumed to be bimodally Gaussian distributed, thus modeling the distribution of both power generators and consumers:
 for the stable operation of power systems these two quantities must be in balance. Since synchronization has to be ensured for a perfectly working power grid,  we investigate the stability of the desired synchronized state. We solve this problem numerically for a population of $N$ rotators regardless of the level of quenched disorder present in the topology.  
 We obtain stable and unstable solutions for different initial phase conditions, and we propose how to control unstable solutions, for sufficiently large coupling strength, 
 such that they are stabilized for any initial phase. Finally, we examine a random Erd{\"o}s-Renyi network under the impact of white Gaussian noise, which is an essential ingredient for power grids in view of increasing renewable energy sources.

\end{abstract}

\pacs{05.45.Xt, 87.18.Sn, 89.75.-k}
\keywords{nonlinear complex networks, power grids, synchronization, stability analysis, control}

\maketitle


\textbf{The goal of this paper is to investigate complex dynamic systems which can model high-voltage power grids with renewable, fluctuating energy sources. 
For this purpose we use the Kuramoto model with inertia to model the network of power generators and consumers. In particular, we analyse the synchronization transition of 
networks of $N$ phase oscillators with inertia (rotators) whose natural frequencies are bimodally distributed, corresponding to the distribution of generator and consumer power. 
Moreover we take into account random Erd\"{o}s-Renyi networks and Gaussian white noise in order to mimic the topological disorder and temporal fluctuations, respectively, characteristics of electric power grids.
The modification of the Kuramoto model by an additional inertial term was firstly reported and investigated by Tanaka, Lichtenberg, and Oishi \cite{TAN97, TAN97a}, who were inspired
by Ermentrout \cite{ERM91} when choosing a phase oscillator model able to show a transition to synchronization via frequency adaptation instead of the usual phase-locking mechanism. That work specifically focussed on a phase oscillator model able to mimic the synchronization mechanism observed in the firefly \textit{Pteroptix malaccae} \cite{ERM91}. 
Recently the model has been used to investigate the self-synchronization emerging in disordered arrays of underdamped Josephson junctions \cite{TRE05}
as well as to show the emergence of explosive synchronization \cite{JI13} in a network of rotators whenever the natural frequency is chosen to be proportional to the node degree.
Nowadays the Kuramoto model with inertia is a standard mathematical model used to study the dynamical behavior of power generators and consumers \cite{FIL08a,FRA12,ROH12,ROH14,OLM14a,NIS15,OLM16,ROH17,GAM17}
since it captures the essential dynamical features of a power grid on coarse scales, but is still simple enough to allow for a comprehensive
understanding of the fundamental properties of power grid dynamics.
}

\section{Introduction}

Within the last century, electrical energy has been produced mainly by power plants based on coal or nuclear power. Nowadays we are witnessing a time of drastic changes in the operation of power grids caused by the necessity to reduce global warming caused by large emission of carbon dioxide gases. Namely, the generating units of a power grid are more and more supplied by natural sources, such as wind parks, photovoltaic arrays and other renewable energy sources. The main question here remains the sustainable and stable operation of power grids, which is of vital importance to our daily lives \cite{MAR08,TUR99}. However, due to the regime shift towards renewable energy sources, three major changes have to be envisaged in modern power grids. The first change is the 
\textit{decentralization}, i.e., the power system represents a distributed network carrying many small units of energy to the consumers instead of large units of energy coming from a few power plants \cite{ACK01}. The second change is a strong \textit{spatial separation} between power generators and consumers. It is evident that power systems based on solar or wind energy should be located in areas 
where such energy is abundant. Finally, the last important change is the increasing fraction of strongly fluctuating power output due to 
renewable energy sources, which are strongly dependent on weather conditions \cite{MIL13,HEI10,HEI11,ANV16,ANV17,SCH17i,SCH18c}.

The investigation of power grid systems has been recently addressed from a nonlinear dynamics point of view, using the Kuramoto phase oscillator model with inertia \cite{FIL08a,FRA12,GAM17,ROH12,ROH14,OLM14a,NIS15,OLM16,ROH17,TUM18,MEH18,TAH19}. This model represents an extended version of the standard Kuramoto model; such an extension has been developed by Tanaka et al \cite{TAN97,TAN97a} 
by including an additional term that takes into account the frequency dynamics. Oscillators are able to reach frequency synchronization by adapting themselves to some intrinsic collective frequency. In particular power grids tend to synchronize their frequencies to the standard ac power frequency $\Omega = 50$ Hz (or $60$ Hz in some countries).
The description of high-voltage power grids in terms of the extended Kuramoto model has been first proposed by Filatrella et al \cite{FIL08a} who distinguished the power generated by power sources 
($P^i_{source}>0$) from the power consumed by passive machines or loads ($P^i_{cons} < 0$). Such a power separation implies using a bimodal distribution of the power (corresponding to the dimensionless natural frequencies in the dimensionless Kuramoto model) within the network. 
Although this is a very important feature of the model, most of the previous studies consider either a unimodal frequency distribution \cite{ACE00, OLM14a} or $\delta$-function shaped bimodal distributions \cite{ACE00,ROH12,ROH14}. 
In our work we use a bimodal Gaussian distribution of frequencies, which models consumed and generated power in a more realistic way.
Moreover, from the topological viewpoint we focus on randomly diluted networks, giving rise to a more realistic description of power grids than all-to-all coupled networks, since real-world systems are characterized by low connectivity per node. 

Stable operation of power grids is characterized by maintaining a synchronous state of the entire network. The present paper is devoted to the stability analysis of a population of $N$ 
Kuramoto oscillators with inertia (rotators) which are randomly connected. A similar stability analysis was performed by Mirollo et al \cite{MIR05} for a network of classical Kuramoto oscillators 
(without inertia). After presenting the model (Sect. II) and discussing the onset of synchronization in a random network (Sect. III), 
we analytically establish the criteria for stability of the synchronized state (Sect. IV) and we solve numerically the dynamics of $N$ coupled rotators by using the Levenberg-Marquard 
algorithm \cite{LEV44,MAR63} (Sect. V). This allows us to obtain, for all $N$ rotators, a set of initial phases for which a frequency synchronized solution evolves. 
Moreover we determine the minimum coupling strength for which frequency synchronization is still possible and we derive both stable and unstable solutions, as illustrated in Sect. VI, 
where solutions are shown as the spatio-temporal evolution of state variables. Unstable frequency synchronized solutions obtained for initial phase differences which are quite distinct from 
zero can be stabilized by applying a suitable control method, as shown in Sect. VII. Thus control enables us to realize a stable frequency synchronized solution even if this does not exist in 
the uncontrolled system for too small coupling constant or non-zero initial phase differences.  Thus control enables us to realize a stable frequency synchronized solution even if this does 
not exist in the uncontrolled system for too small coupling constant or non-zero initial phase differences.  Finally, in Sect. VIII, we add temporal random fluctuations and consider the 
interplay of Gaussian white noise and spatial disorder due to diluted connectivities. We examine how the frequency synchronized solution changes with respect to the deterministic case. 
A discussion of all results is finally given in Sect. IX.

\section{Model}
\label{sec:model}

The investigated system consists of a population of $i=1,...,N$ coupled Kuramoto oscillators with inertia and reads
\begin{equation}\label{EQ:1}
m \ddot\theta_i + \dot\theta_i = \Omega_i + \frac{K}{N_i} \sum\limits_{j=1}^N A_{ij} \sin\left(\theta_j - \theta_i\right),
\end{equation}
where $\theta_i$ and $\dot\theta_i$ are the instantaneous phase and frequency, respectively, of the oscillator $i$. In terms of the power grid, $\dot\theta_i$ represents the frequency
deviation of the $i$-th oscillator with respect to the standard grid frequency ($50$ or $60$ Hz). 
The parameter $m > 0$ indicates the inertial mass of generators and loads that plays a fundamental role in determining the hysteretic transition to synchronization. 
$K > 0$ is the coupling constant of the network, which measures the strength of the connectivities among the oscillators. 
In terms of power grid systems, $K$ is equivalent to the transmission line capacities between loads and generators. Increasing coupling strength enhances the synchronization
of elements with heterogeneous natural frequencies. $A$ is the connectivity matrix, whose entries $A_{ij}$ can be either one, provided there is 
a link between the nodes $i$ and $j$, or zero, if the link is absent. From the topological viewpoint a power grid network is an undirected, symmetric graph, i.e., $A_{ij} = A_{ji}$. 
$N_i$ is the node degree of the $i$-th element, thus denoting the number of the links outcoming from this node. Finally $\Omega_i$ represents the natural frequency 
of the oscillator $i$, whose value is chosen in accordance with a bimodal Gaussian distribution 
\begin{equation}\label{EQ:2}
g(\Omega) = \frac{1}{2\sqrt{2 \pi}} \left[ e^{-\frac{(\Omega - \Omega_0)^2}{2}} + e^{-\frac{(\Omega + \Omega_0)^2}{2}} \right].
\end{equation}
In particular $g(\Omega)$ is the superposition of two Gaussians with unit standard deviation, whose peaks are located at  $\Omega_0$ and $-\Omega_0$. 
Thus, the distance between the peaks is $2\Omega_0$.  In the following we will assume almost non-overlapping Gaussians, i.e., we choose  $\Omega_0=2$.

The physical motivation for chosing a bimodal distribution comes directly from Filatrella et al. \cite{FIL08a} : according to their work, each element of the power grid network 
either generates ($P^i_{source} > 0$) or consumes ($P^i_{cons} < 0$) power. Thus, one should distinguish two kinds of oscillators: the \textit{sources} which deliver electrical power, 
and the \textit{consumers} which consume this power. Hence, the electrical power distribution of all oscillators should be bimodal, with a maximum at $P^i_{source} > 0$ and one at $P^i_{cons} < 0$. 
In the dimensionless Kuramoto model with inertia given by eq. \eqref{EQ:1} this corresponds to a bimodal frequency distribution of the $\Omega_i$ that we assume to be given by a superposition of two 
Gaussians with peaks at corresponding positive and negative frequencies. Thus the necessary condition for the existence of the steady state is that the sum of the generated 
power equals the sum of the consumed power in order for the energy to be conserved. 

The phase ordering of the power grid is measured by the complex order parameter
\begin{equation}\label{EQ:3}
r(t) e^{i\phi(t)} = \frac{1}{N} \sum\limits_{j=1}^N e^{i\theta_j},
\end{equation}
where its modulus $ r(t) \in \left [ 0 , 1 \right ] $ and argument $ \phi(t) $ indicate the degree of synchrony and mean phase angle, respectively. 
In the following we will denote $ r(t) $ as \textit{global order parameter}. In the continuum limit an asynchronous state is characterized by $r \approx 0$, while $r=1$ 
corresponds to full phase synchronization. Intermediate values of $r$ correspond to states with partial or cluster synchronization. 

Throughout this study we will mainly consider Erd\"{o}s-Renyi networks , i.e., the graph is constructed by connecting nodes randomly. This topology turns out to be more realistic in comparison to 
a globally coupled network, since power grid networks are characterized by only few links per node. We assume a constant node degree $N_i=N_c$ 
and a dilution parameter $p=\frac{N_{c}}{N}$. The latter indicates the ratio of existing links to the number of all possible links. These so-called \textit{diluted} networks 
are thus obtained by considering random realizations of the coupling matrix $A$, keeping  the connectivity matrix symmetric.

\section{Diluted networks: the onset of synchronization}

First, we explore the transition to synchronization for a randomly coupled set of power suppliers and consumers described by eq. \eqref{EQ:1}. The random network we investigate is characterized 
by dilution parameter $p=0.20$, thus indicating that each node is randomly connected to $20 \%$ of all possible $N-1$ nodes. 
A typical synchronization transition profile is illustrated in Fig. 1 (a), where we show the time-average global order parameter obtained by sweeping up and down adiabatically 
the connectivity strength $K$, following two different protocols as described in \cite{TAN97,OLM14a, OLM16,TUM18}. In particular \textit{protocol (I)} denotes the up-sweep: 
the system's state variables $\left \{ \theta_i \right \}$ and $\left \{\dot{\theta}_i \right \}=\left \{ \omega_i \right \}$ are randomly initialized in absence of coupling; 
then the coupling strength is increased in steps of $\Delta K = 0.5$ until the maximum coupling $K_M$ is reached (for $p=0.2$ we choose $K_M=60$). 
Note that the global order parameter increases as the elements become more strongly connected. Finally it reaches the maximum value $\bar{r} \approx 1$ as synchronization is achieved, 
which corresponds to the maximum coupling $K_M$. At each step the initial conditions for phases and frequencies correspond to the final conditions obtained for the previous $K$ value.
By \textit{protocol (II)} we denote the reverse procedure: this time the initial state corresponds to the synchronized system at $K=K_M$, while the coupling is adiabatically decreased in steps 
$\Delta K = 0.5$, until we approach again a completely uncoupled asynchronous system. For both protocols the investigation of the nature of the dynamics emerging at each time step is done by using 
the same procedure: the system is simulated for a transient time $T_R$ followed by an investigation period $T_W$, during which the average values of global order parameter $\bar{r}$, the phase velocities 
$\left \{ \bar{\omega}_i\right \}$ and the maximum natural frequency of the locked oscillators are calculated. 

Now we focus on a more detailed description of the different regimes emerging in the system by varying the coupling strength $K$, see Fig. 1 (a). 
For small coupling constant the system is uncoupled and asynchronous (AS), characterized by a low value of the time-averaged order parameter $\bar{r} \approx 1/\sqrt{N}$ and non-identical average phase 
velocities $\bar{\omega}_i$ for all the elements $i$, see Fig. 1 (b). Increasing the coupling $K$ leads to a rapid jump of the average global order parameter $\bar{r}$ to higher values, i.e. $\bar{r} > 0.1$. 
Here we observe the emergence of one or more clusters of locked oscillators characterized by nodes with the same average phase velocity $\bar{\omega}_i$.
The coexistence of chaotically whirling oscillators with clusters of locked nodes corresponds to a traveling wave (TW) solution (see Fig. 1 (c)) that is observable for $K>K^{TW}$.
A further increase of coupling can cause both the enlargement of the existing clusters of locked oscillators and the collapse of smaller clusters to larger ones, which are usually characterized by 
an average phase velocity $\bar{\omega}_i \approx \pm \Omega_0$. For $K>K^{SW}$ the system continuously approaches the standing wave state (SW), which is characterized by two symmetric clusters 
of locked oscillators drifting with opposite average phase velocities equal to $\bar{\omega}_i \approx \Omega_0$ and $-\Omega_0$ (see Fig. 1 (d)). In the SW regime the system behaves like two independent 
subnetworks each one corresponding to a network with unimodal Gaussian frequency distribution, whose peaks are located respectively at $+ \Omega_0$ (generators) or $-\Omega_0$ (loads). 
The corresponding time-averaged global order parameter equals $\bar{r} \approx 0.5$. 
Finally, for further increase of the coupling $K$, the average global order parameter $\bar{r}$ exhibits a rapid jump to higher values, i.e., $\bar{r} > 0.9$. 
This means that for $K>K^{PS}$, the system reaches a partially (almost completely) synchronized regime. Thereby the two subnetworks that for smaller K behave almost independently, now merge into a 
unique stationary cluster with $\bar{\omega}_i \approx 0$, see Fig. 1 (e). On the other hand, the number of unlocked oscillators is vanishingly small, i.e., $N-N_L \approx 0$. 
Such a rapid change of average global order parameter $\bar{r}$ allows us to identify the onset of synchronization of a network. As we continue increasing $K$
the system smoothly approaches the regime of full synchronization. 

While the transition to synchronization for $K>K^{PS}$ is always detectable, irrespectively of the chosen value of the dilution parameter $p$,
the standing wave regime is not always detectable as the dilution increases and it actually disappears as the network topology becomes highly diluted, e.g., for $p<0.05$. 
In particular, as the randomness increases, it becomes more and more difficult for the system to reach such states as many elements will have different subgraphs of connected nodes
with a variable percentage of nodes belonging to the same native class or to the opposite one, where the classes identify the oscillators with positive or negative natural frequencies,
respectively. Therefore the separation in two subnetworks with positive and negative classes, leading to a configuration similar to the one shown in Fig. 1 (d) is hardly achieved.
Finally the disappearence of SW turns into a decrease of the critical value $K^{PS}$, as previously reported \cite{TUM18}.

If we analyse the system in accordance with protocol (II), the syncronous state survives for a large $K$ interval until it collapses towards asynchronicity at $K < K^{DS}$, 
where DS denotes desynchronization. Note that there is a considerable difference between the critical coupling values required to synchronize or desynchronize the system and $K^{PS}>K^{DS}$.
In other words, the system needs a stronger coupling to become synchronized with respect to the desynchronizing value and once it is synchronized, due to inertia, it hardly leaves
this regime. The transition to synchronization is therefore \textit{hysteretic} and the size of the hysteresis loop  $K^{PS}-K^{DS}$ depends on the inertia $m$, 
$K^{PS}$ is strongly affected by the dependence on $m$ \cite{OLM14a, TUM18}.

\begin{figure}[]
\begin{minipage}[h]{0.99\linewidth}
\center{\includegraphics[width=\linewidth]{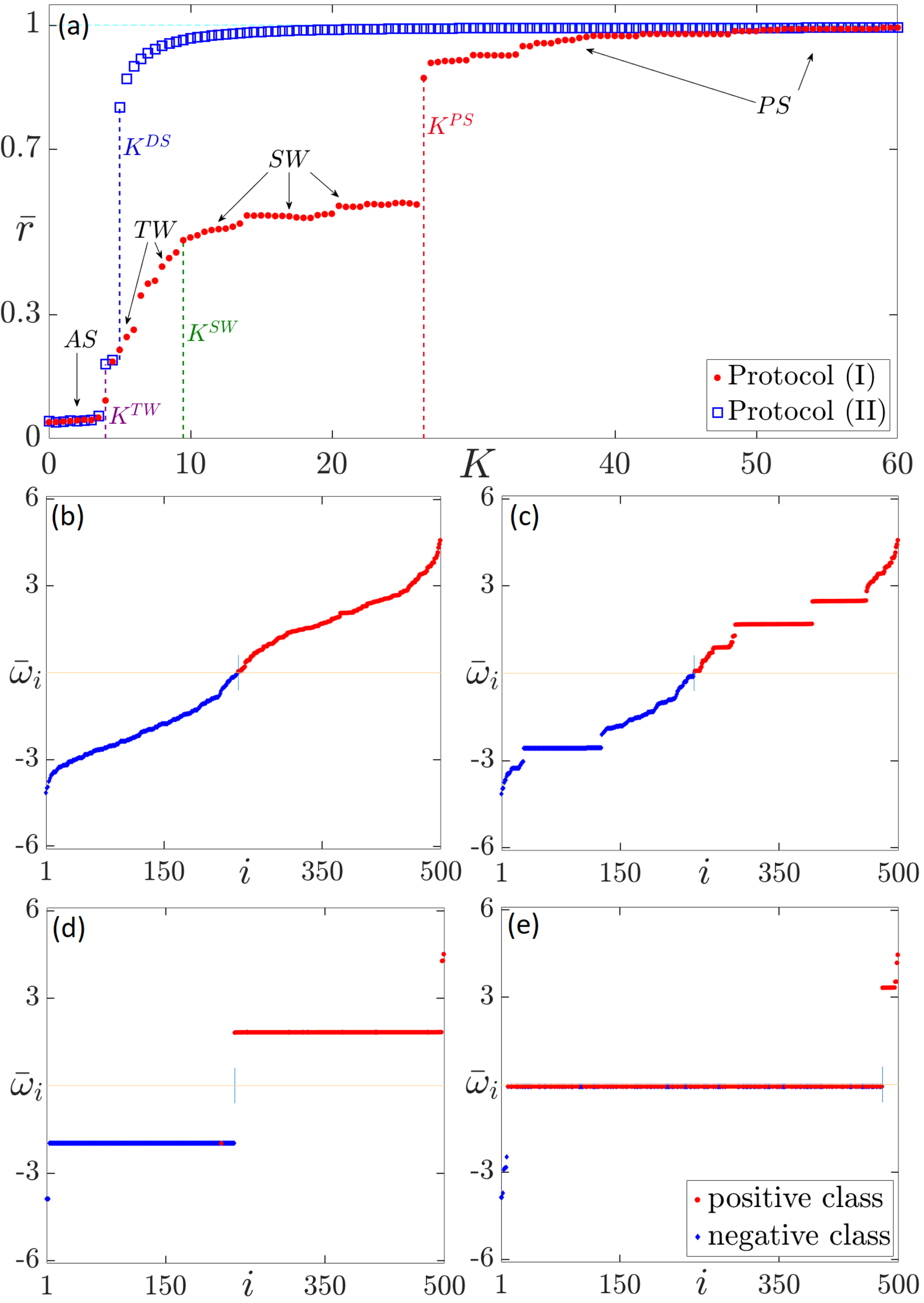}}
\caption{(a): Time-averaged global order parameter $\bar{r}$ as a function of coupling constant $K$ for two series of simulations, obtained by following the protocol (I) 
(upsweep, red filled circles) and (II) (downsweep, blue empty squares) for a diluted network. The vertical dotted lines indicate the critical values of coupling $K$ for traveling waves ($K^{TW}$, purple), standing waves ($K^{SW}$, green), 
partial synchronization ($K^{PS}$, red) and the value at which desynchronization occurs ($K^{DS}$, blue). Average phase velocity $\bar{\omega}_i$ as a function of node $i$ for (b) $K = 2$, $\bar{r} = 0.043$ (asynchronous state); 
(c) $K = 5$, $\bar{r}=0.217$ (traveling wave); (d) $K = 25$, $\bar{r} = 0.567$ (standing wave); (e) $K=33$, $\bar{r}=0.928$ (partial synchronization). The nodes are labeled such that the average phase velocities $\bar{\omega}_i$ are sorted from low to high values. Positive (red) and negative (blue) classes refer to positive and negative natural frequency distributions, respectively.
Parameters: $m=6$, $p=0.20$, $N=500$, $T_R = 4000$, $T_W = 200$ ($\alpha=1/6$, $\sigma=K/(pNm)=K/600$).}
\label{FIG:1}
\end{minipage}
\end{figure}

\section{Synchronous solution}

Synchronization is a mandatory regime when stable operation of power grids is required. Therefore, determining the stability of synchronous states is one of the central goals of the present study. 
In particular we aim to investigate the stability of the synchronous solution emerging in a power grid network by calculating the maximum Lyapunov exponent $\mu_{max}$, whose sign 
will be the main criterion for determining the synchronization stability.
If we re-write eq. \eqref{EQ:1} in terms of two dynamical variables, i. e., phase $\theta_i$ and frequency $\omega_i$, we obtain the following $2N$-dimensional first-order system

\begin{equation}\label{EQ:4}
\begin{aligned}
\dot \theta_i &= \omega_i \\
\dot \omega_i &= \alpha \left( \Omega_i - \omega_i \right ) + \sigma \sum\limits_{j=1}^N A_{ij} \sin(\theta_j - \theta_i), 
\end{aligned}
\end{equation}
where $ \alpha = \frac{1}{m} $ and $\sigma = \frac{K}{N_c m}$.

Phase synchronization implies for all the phases $\theta_1 = \theta_2 = \dotsc = \theta_c $. If we denote the corresponding frequency as $\omega_c$, 
since the coupling term $\sigma \sum\limits_{j=1}^N A_{ij} \sin(\theta_j - \theta_i) = 0$, we obtain from eq. \eqref{EQ:4} 

\begin{equation}\label{EQ:5}
\dot \omega_c = \alpha \left( \Omega_i - \omega_c \right ) \Rightarrow \Omega_i = {\alpha}^{-1} \dot \omega_c + \omega_c
\end{equation}
which holds only for the trivial case $\Omega_i=const$, since the right side of equation is not a function of $i$, while we have assumed a bimodal distribution for the natural frequencies.
This suggests that complete \textit{phase synchronization is not achievable} in our system.

Moreover, in case of frequency synchronization: $\dot \theta_1=\dotsc = \dot \theta_n = \omega_c \Rightarrow \theta_i - \theta_j = const$. 
If we rename the coupling term as $\chi_i = \sigma \sum\limits_{j=1}^N A_{ij} \sin(\theta_j - \theta_i) $, from eq. \eqref{EQ:4} we get
\begin{equation}\label{EQ:6}
\dot \omega_c = \alpha \Omega_i - \alpha \omega_c + \chi_i.
\end{equation} 

Note that the variables $\dot \omega_c $ and $\omega_c$ do not depend on index $i$. Hence we can define a constant $C_0$ such that
\begin{equation}
\nonumber
\dot \omega_c + \alpha \omega_c = \alpha \Omega_1 + \chi_1 = \dotsc = \alpha \Omega_n + \chi_n = C_0.
\end{equation}
By using the previous definition of $C_0$, we get a system of differential equations for the frequency synchronized solution $\dot \omega_c = C_0 - \alpha \omega_c$.
If we assume to be in a frequency-synchronized regime ($\dot \omega_c = 0$), thus allowing us to identify
\begin{equation}\label{EQ:8}
\omega_c = \frac{1}{\alpha} C_0.
\end{equation}
Finally the constant $C_0$ can be calculated by summing eq. \eqref{EQ:6} over all the nodes $i=1,\ldots,N$
\begin{equation}
\label{EQ:9}
\sum\limits_{i=1}^N \left [ \alpha \Omega_i - \alpha \omega_c + \sigma \sum\limits_{j=1}^N A_{ij} \sin(\theta_i - \theta_j) = 0 \right ]. \\
\end{equation}
Since the chosen network is a symmetric undirected graph, the term $\sum\limits_{i=1}^N \sum\limits_{j=1}^N A_{ij} \sin(\theta_i - \theta_j) = 0$
and Eq.~\ref{EQ:9} reduces to
\begin{equation}
\nonumber
\sum\limits_{i=1}^N \alpha \Omega_i - \sum\limits_{i=1}^N \alpha \omega_c = 0 \Rightarrow \sum\limits_{i=1}^N \alpha \Omega_i - N \alpha \omega_c = 0. 
\end{equation}
Thus it follows that for a finite system the value at which the frequency synchronizes is the arithmetic mean of all the natural frequencies
\begin{equation}\label{EQ:11} 
\omega_c = \frac{1}{N}\sum\limits_{i=1}^N \Omega_i.
\end{equation}
For large network size we expect this value to be close to $0$. In the present case the arithmetic mean will be substituted with the expectation value, since natural frequencies are distributed according to a bimodal Gaussian distribution as shown in eq. \eqref{EQ:2}.

\section{Stability analysis of frequency synchronized solution}

The phase evolution of the system in case of full frequency synchronization is given by
\begin{equation}\label{EQ:12}
\theta^t_{i} = \theta^0_{i} + \omega_c t,
\end{equation}
where $\theta^0_{i}$ denotes the initial phase of the oscillator $i$ at time $t=0$. Using the equality $\sin(\theta^t_{i} - \theta^t_{j}) = 
\sin(\theta^0_{i} - \theta^0_{i})$, one can write 
\begin{equation}
\nonumber
\alpha \Omega_i - \alpha \omega_c + \sigma \sum\limits_{j=1}^N A_{ij}\sin(\theta^0_{j} - \theta^0_{i}) = 0.
\end{equation}
If we rename $\tilde \omega_i = \alpha (\Omega_i - \omega_c)$, the previous equation reads
\begin{equation}\label{EQ:14}
\sigma \sum\limits_{j=1}^N A_{ij} \sin(\theta^0_{j} - \theta^0_{i}) = - \tilde \omega_i,
\end{equation}
which turns out to be fundamental in order to determine the values of initial phases $\theta^0_{i}$ necessary to obtain a frequency synchronized solution.

In order to determine the linear stability, consider the evolution of the system \eqref{EQ:4} subject to small perturbations around the desired (frequency synchronized) solution, 
i. e., $\theta_i = \theta^t_{i} + \delta{\theta}_i $, $\omega_i = \omega_{c} + \delta{\omega}_i $
\begin{equation}\label{EQ:15}
\begin{aligned}
\delta\dot{\theta}_i &= \delta\omega_i  \\
\delta\dot{\omega}_i&= - \alpha \delta\omega_i + \sigma \sum\limits_{j=1}^N A_{ij} \cos(\theta^0_{j} - \theta^0_{i}) (\delta\theta_j - \delta\theta_i).
\end{aligned}
\end{equation}
The system of $2N$ ordinary differential equations can be rewritten in a block matrix form

\begin{eqnarray}\label{EQ:17}
 \begin{pmatrix}
    \delta \dot{\theta} \\
    \delta \dot{\omega} \\
\end{pmatrix} 
     =
    \begin{pmatrix}
     \bf{0} & \mathbb{I} \\
     \sigma\bf{M} & -\alpha \mathbb{I} \\
\end{pmatrix}
 \begin{pmatrix}
    \delta\theta \\
    \delta\omega \\
\end{pmatrix} ,
\end{eqnarray}
where $\delta \theta \equiv(\delta \theta_1,...,\delta \theta_N)^T$,  $\delta \omega \equiv(\delta \omega_1,...,\delta \omega_N)^T$, $\mathbb{I}$ is the $N \times N$ unit matrix, and \textbf{M} represents the $N \times N$ Laplacian matrix of a weighted undirected graph
\begin{eqnarray*}
\footnotesize{\bf{M} 
     =
    \begin{pmatrix}
     - \sum\limits_j A_{1j} \cos(\theta^0_{j} - \theta^0_{1}) & \cdots &   A_{1N} \cos(\theta^0_{1} - \theta^0_{N})   \\
    \cdots &  \cdots &  \cdots  \\
      A_{N1} \cos(\theta^0_{N} - \theta^0_{1}) & \cdots &  - \sum\limits_j A_{Nj} \cos(\theta^0_{j} - \theta^0_{N})   \\
\end{pmatrix}}. 
\end{eqnarray*}
The stability of the frequency synchronized solution can be analyzed by solving the eigenvalue problem of the $2N \times 2N$ matrix appearing in eq. \eqref{EQ:17}
\begin{eqnarray}\label{EQ:18}
\textrm{det}(\mathbf{G} - \lambda \mathbb{I}) \equiv
\begin{vmatrix}
-\lambda \mathbb{I} & \mathbb{I} \\
\sigma\bf{M} & (-\lambda - \alpha)\mathbb{I} \\ 
\end{vmatrix}.
\end{eqnarray}
By using \textit{Schur's complement} we obtain \ $\left |  \mathbf{G} \right | = \left |  G_{11} \right | \left |  \frac{\mathbf{G}}{G_{11}} \right | = 
\left |  G_{11} \right | \left | G_{22} - G_{21} {G_{11}}^{-1} G_{12} \right |$. Thus, we are able to derive an expression for $\textrm{det}(\mathbf{G} - \lambda \mathbb{I}) = 
\left | -\lambda \mathbb{I} \right | \left | (-\lambda - \alpha) \mathbb{I} - \sigma \bf{M}({-\lambda \mathbb{I}})^{-1}\mathbb{I}  \right | = {-\lambda}^{N} \left | (-\lambda - \alpha)\mathbb{I} 
+ {\lambda}^{-1} \sigma \bf{M} \right | $ and finally obtain
\begin{equation}\label{EQ:19}
 \left | (\lambda^2 + \lambda \alpha)\mathbb{I} - \sigma \bf{M} \right |=0.
\end{equation}
If we denote the eigenvalues of the matrix \textbf{M} by ${\mu}$, i.e., $\left | \mu \mathbb{I} - \bf{M} \right | = 0$, then  we have to solve quadratic equations of the type 
\begin{equation}
\nonumber
\lambda^{2} + \lambda \alpha - \sigma \mu = 0
\end{equation}
in order to determine the eigenvalues of the matrix \textbf{G} defined in eq. \eqref{EQ:18}. The eigenvalues are given by
\begin{equation}\label{EQ:22}
\lambda =  \frac{-\alpha \pm \sqrt{\alpha^2 + 4\mu \sigma}}{2}
\end{equation}
and depending on the properties of \textbf{M} the following holds:
\begin{enumerate}[label=(\roman*)] 
\item $\exists \mu > 0 \Leftrightarrow \exists \lambda > 0$. 
\item If the matrix $\bf{M}$ is stable, then $\bf{G}$ is also stable. 
\item If  $\cos(\theta^0_{j} - \theta^0_{i}) > 0$ ($ \left | \theta^0_{j} - \theta^0_{i} \right | < \frac{\pi}{2}$), then $\bf{M}$ is a \textit{diagonally dominant matrix}. 
This means $\left | M_{ii} \right |  \ge \sum\limits_{j \not= i} \left | M_{ij} \right |$).
\end{enumerate}
Thus we can conclude that it is not possible to find an unstable solution in the neighborhood of $\left\lbrace \theta^0_{i}\right\rbrace =0$.  Furthermore, we know that any matrix $\bf{R}$ is positive definite 
if $R_{ii} > 0 $, thus $\lambda(\bf{R}) \ge 0 $. In our case, matrix $-\bf{M}$ satisfies this conditions, thus $\mu \le 0$. In accordance with (ii), it comes straightforward that the whole system \eqref{EQ:17} is stable with respect to small perturbations.

\section{Numerical solution of eigenvalue problem}

In this section we explicitly solve to the eigenvalue problem stated in Sec. V. In order to perform the stability analysis of a frequency synchronized solution, 
we need to find the phases which satisfy the condition expressed in eq. \eqref{EQ:14}. This means solving an $N$ dimensional system with $N$ unknown variables, namely, the phases $\theta_i$. 
This system may have several solutions, the number of which depends on the system parameters. For instance, low values of the coupling strength $K$ might not admit any solution at all. 
Nevertheless, for a proper parameter choice there exist phases $\theta^\ast_{i}$ satisfying eq. \eqref{EQ:14}, thus describing trajectories a 
frequency synchronized system follows. Once the set of phases $\theta^\ast_{i}$ is found, we will insert them into the Eqs.~\eqref{EQ:17} to solve the eigenvalue problem 
of the Laplacian matrix \textbf{M}. This, finally, will enable us to characterize the stability of the frequency-synchronized solution. 

We are looking for a set of initial phases $ \theta^\ast_{i}$ such that the following equation holds:
\begin{equation}\label{EQ:23}
\textrm{F}_{i}(\theta^\ast) = \tilde \omega_i + \sigma \sum\limits_{j=1}^N A_{ij} \sin(\theta^\ast_j - \theta^\ast_i) = 0. 
\end{equation}

This multidimensional problem can be solved numerically by using the \textit{Levenberg-Marquardt algorithm}, which represents a combination of a Gauss-Newton algorithm and the method of gradient descent. 
The algorithm minimizes, with an iterative procedure, the sum $\sum |F_i|^2$, given an initial guess for $\theta^0_i$. 
After a certain number of iterations the algorithm converges to a local minimum near the initial guess for the phases $\theta^0_i$. 
However, in order to guarantee that the obtained solution is a true solution for the system of equations \eqref{EQ:23}, it should fulfill the condition $\sum |F_i|^2 \equiv 0$. 
The trivial choice for the initial conditions $\left\lbrace \theta^0_i\right\rbrace \approx 0$ gives rise to a suitable solution only if the coupling $K$ is sufficiently strong. 
A heuristic explanation for this is that in the limit $K \to \infty$ the choice of phases $\theta^\ast_i \equiv 0$ is always the solution, independently of $\tilde\omega_i$ and coupling matrix \textbf{A}. 
On the other hand, it is always possible, especially for large $K$, to find solutions which are quite distinct from the solution obtained at initial zero phases. Recalling that 
$ \left | \theta^0_{j} - \theta^0_{i} \right | < \frac{\pi}{2}$ must be satisfied in order to ensure that the matrix \textbf{M} is diagonally dominant, these new solutions will be unstable.
Finally, once a set of phases $ \theta^\ast_i$ which minimizes the function is found, it is possible to solve the eigenvalue problem for the matrix $\mathbf{M}$. 
The eigenvalue with maximum real part determines the stability of the state, while $\mu = 0$, which is always present in the Laplacian corresponding to the invariance of the model 
under uniform phase shift, will not be considered in the following.

Our system is characterized by a sparse network of Kuramoto oscillators with inertia, whose parameters $m=6$, $p=0.20$, $N=500$ are the same as in Fig. 1. By means of the Levenberg-Marquardt algorithm, we are able to obtain a set of phases $\theta^\ast_i$,  which fulfills the condition for frequency synchronization Eq.~\eqref{EQ:23}. 
However, as previously discussed, it is not always possible to find a solution (stable or unstable) to our system for arbitrary $K$, therefore we estimate the critical coupling $K_c$, below which the nonlinear system \eqref{EQ:23} has no solutions. Without loss of generality we choose as initial guess for the phases $\theta^0_i\equiv 0$ which is close to the true phases solving Eq.~\eqref{EQ:23}.
The results are illustrated in Fig. 2. In particular we can observe that no solution can be found for $K<K_c=5.8$ while we can always find a solution for $K > K_c$. 
If $K$ is only slightly above the critical value $K_c$, then few solutions are admittable for phases $\theta^\ast_i$. For $K$ slightly above$K_c$ the coupling is not sufficiently 
strong to suppress the phase differences $ \left | \theta^0_{j} - \theta^0_{i} \right |$, and for this reason the global order parameter $r \approx 0.9$. 
The corresponding maximum Lyapunov exponent (disregarding $\lambda=0$) is negative. 
By further increasing $K$ a stable solution is obtained for smaller phase differences $ \left | \theta^0_{j} - \theta^0_{i} \right |$, which contribute to an increase of the global order parameter $r$. 
The corresponding maximum eigenvalue $\mu_{max}$ of $\textbf{M}$ decreases accordingly (inset of Fig.2). The eigenvalue $\lambda$ determining the stability of the frequency-synchronized solution is given by Eq.~(15) and has the real part Re($\lambda)=-\alpha/2=-0.083$ for $4|\mu| \sigma>\alpha^2$ which holds for all $K > K_c = 5.8$ where the frequency-synchronized solution exists.

It is remarkable that the stable solution obtained by iterating the algorithm with initial phases set to zero, coincides with the simulations obtained by performing protocol (II) 
(at least outside the shaded area). This suggests that the hysteretic loop observed in Fig. 1, due to the presence of the inertial term, 
strongly depends on the initial conditions for the phases. It follows that by choosing an appropriate set of initial phases $\theta^{0}_i$, the power grid can reach synchronization faster without passing through intermediate states (i.e., traveling and standing waves) as in protocol (I).
\begin{figure}[]
\begin{minipage}[h]{0.99\linewidth}
\center{\includegraphics[width=\linewidth]{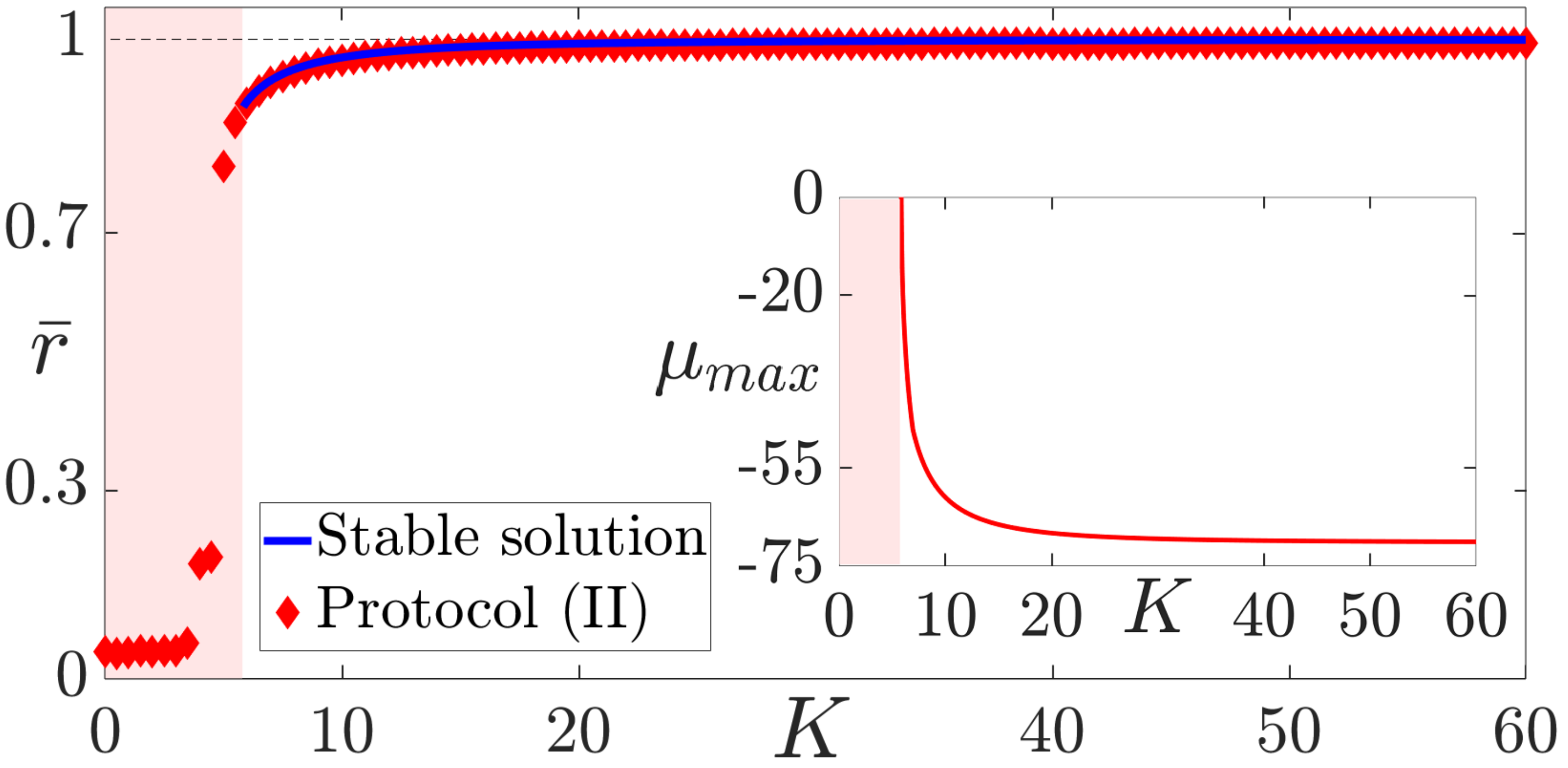}}
\end{minipage}
\caption{Average global order parameter $\bar{r}$ as a function of coupling strength $K$. The shaded red area indicates the region in which no frequency-synchronized solutions can be found. 
The blue solid line refers to the stable solution, whereas the sequence of simulations obtained by running Protocol (II) is denoted by red diamonds. 
Inset: maximum eigenvalue $\mu_{max}$ as a function of coupling strength $K$. Other parameters as in Fig. 1.}
\label{FIG:2}
\end{figure}

The solution of the eigenvalue problem, related to a specific set of phases $\theta^\ast_{i}$, gives rise both to stable and unstable solutions. 
An example of this is illustrated in Fig. 3, where stable (unstable) solutions are reported in panels a,b (c,d). In particular panel (a) depicts the spatio-temporal evolution of the phases $\theta_i$ for a stable solution (Re($\lambda_{max})= -0.083$): the initial set of phases $\theta^\ast_{i}$ is approximately equal to zero while, as time changes, 
$\theta_i$ change at equal rates for all the nodes, thus indicating that they move with the same constant phase velocity $\omega_c$, as confirmed by the evenly distributed greenish color in panel (b). Since the natural frequencies $\Omega_i$ are randomly distributed according to a bimodal Gaussian distribution with opposite means $\Omega_0 = \pm 2$, the arithmetic mean $\omega_c$ given in eq. \eqref{EQ:11} 
is close to $0$.  

A different scenario arises for the unstable solution, see the bottom panels of Fig. 3. Initially, all the phases $\theta^\ast_{i}$ obtained by solving 
eq. \eqref{EQ:23} are uniformly distributed on a limit cycle. When the system starts evolving, the phases $\theta_i$ evolve non-trivially in time and they change at equal rates up to the time $t \approx 9$. 
Afterwards, the system starts oscillating until a new solution is reached. A confirmation of this behavior can be found by analyzing the temporal evolution of the corresponding 
frequencies $\omega_{i}$ shown in panel (d). Here we see that the frequencies lose their constant value at $t \approx 9$, corresponding to frequency synchronization death.  
For different sets of initial conditions, while keeping the same coupling constant $K$, it is also possible to observe cases where the system leaves the frequency-synchronized solution even earlier with larger Re($\lambda_{max}$). Finally, for increasing coupling $K$ the system approaches another state.

\begin{figure}[]
\begin{minipage}[h]{0.99\linewidth}
\center{\includegraphics[width=\linewidth]{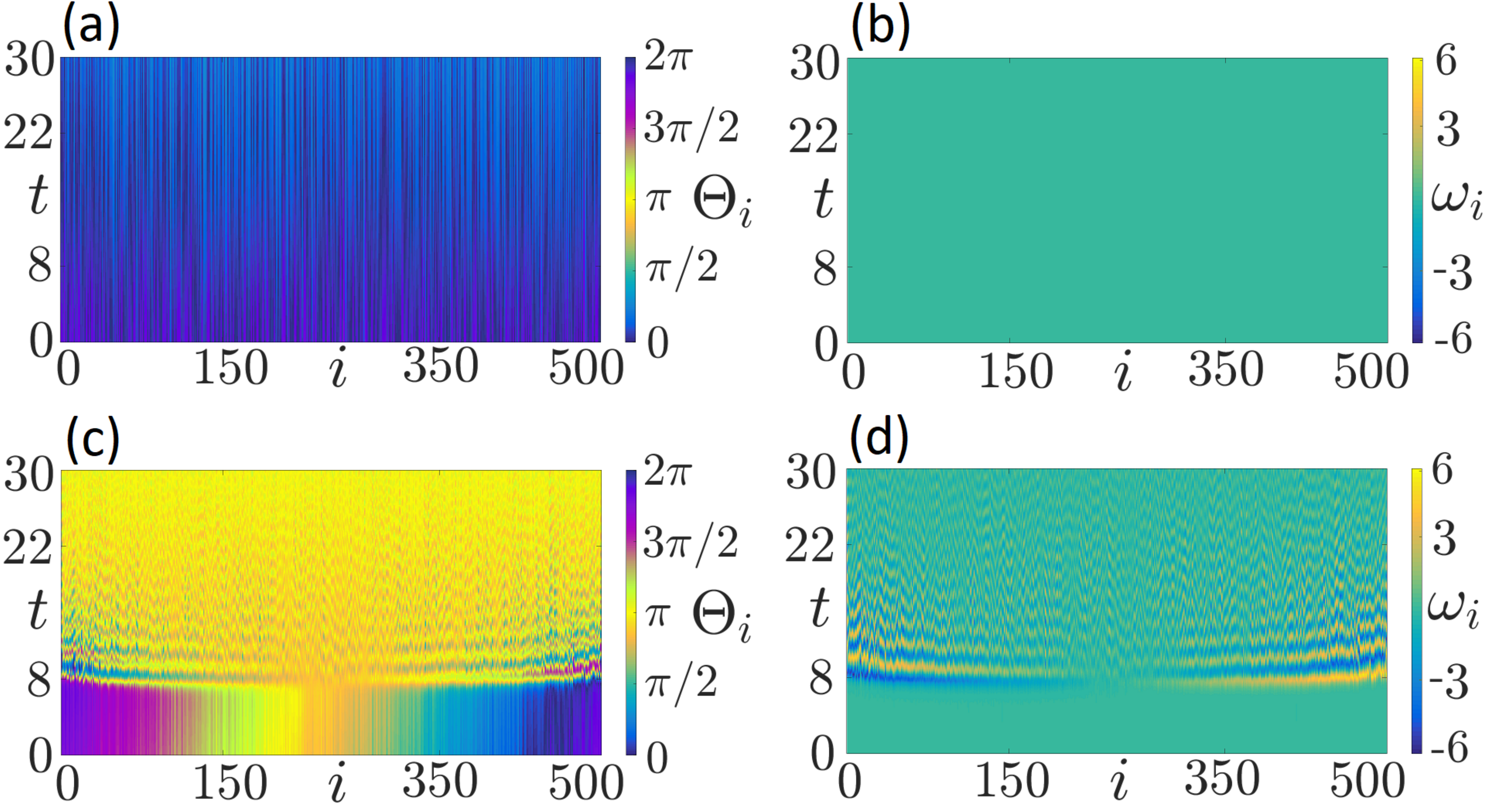}}
\end{minipage}
\caption{Spatio-temporal evolution of phases $\theta_{i}$ and frequencies $\omega_{i}$, which satisfy the condition for frequency synchronized solution. Stable solution: (a) phases; (b) frequencies; parameters: $Re(\lambda_{max}) = -0.083$, $K=10$. Unstable solution: (c) phases; (d) frequencies; parameters: 
$\lambda_{max}=2.41$, $K=70$. Other parameters: as on Fig. 1.}
\label{FIG:3}
\end{figure}

\section{Control of unstable states}

In this section we aim to control the stability of the solution satisfying eq. \eqref{EQ:23}. Namely, we want to stabilize frequency-synchronized solutions obtained for sets of initial phases which violate the condition for diagonal dominance of the matrix $\mathbf{M}$, i.e., for sets of phases whose differences $ \left | \theta^0_{j} - \theta^0_{i} \right | \geq \frac{\pi}{2}$. 
For this purpose we introduce a control term $u_i$ into the original system \eqref{EQ:4}
\begin{equation}\label{EQ:24}
\begin{aligned}
\dot \theta_i &= \omega_i \\
\dot \omega_i &= \alpha \Omega_i - \alpha \omega_i + \sigma \sum\limits_{j=1}^N A_{ij} \sin(\theta_j - \theta_i) + u_i, 
\end{aligned}
\end{equation}
The linearized Eqs.~\eqref{EQ:17} will change accordingly
\begin{eqnarray}\label{EQ:25}
 \begin{pmatrix}
    \delta \dot{\theta} \\
    \delta \dot{\omega} \\
\end{pmatrix} 
     = \mathbf{G}
    \begin{pmatrix}
     \delta \theta \\
     \delta \omega \\
\end{pmatrix} + \mathbf{B} u,
\end{eqnarray}
where $\mathbf{G} \in R^{2N \times 2N}$ and $\mathbf{B} =  \begin{pmatrix}
   \mathbb{O}_{N}  \\
   \mathbb{I}_{N}
\end{pmatrix}  \in R^{2N \times N}$.

In particular the control term $u$ can be chosen as a feedback control loop such that
\begin{eqnarray}\label{EQ:26}
 u = -\mathbf{C} \begin{pmatrix} \delta\theta  \\ \delta\omega \end{pmatrix} 
\end{eqnarray}
where $ \mathbf{C} \in R^{N \times 2N}$ is chosen to minimize the following cost functional
\begin{eqnarray}\label{EQ:27}
J(u) = \int_{0}^{\infty} \left | \left| \begin{pmatrix} \delta\theta(t) \\ \delta\omega(t) \end{pmatrix} \right |\right|^{2} + \left || u(t) \right ||^{2} dt.
\end{eqnarray}
This problem is solved via the application of a \textit{linear quadratic regulator} for each set of phases $\theta^\ast_{i}$.
Basically, the regulator chooses the matrix $\mathbf{C}$ such that the eigenvalues for the closed-loop system $\lambda_{ctrl}$ are non-positive when solving the eigenvalue problem for the 
matrix $\mathbf{G} - \mathbf{B}\mathbf{C}$.  Thus the frequency synchronized solution obtained from eq.\eqref{EQ:23} is stabilized for each particular set of chosen phases $\theta^\ast_{i}$,
and, regardless of the initial phase differences $ \left | \theta^0_{j} - \theta^0_{i} \right | $, we are always able to obtain a stable solution. 

The efficiency of the control method is shown in Fig. 4  for two different sets of initial phases $\theta^\ast_{i}$. In particular in panel (a) we use an initial set of phases which consists of approximately $70 \%$ of phases set to  $\theta^\ast_{0} = 0$ and the remaining $30 \%$ set to $\theta^\ast_{0} = \pi$, which does not fulfill the criterion for the diagonal dominance of matrix \textbf{M}, since $ \left | \theta^0_{j} - \theta^0_{i} \right | > \frac{\pi}{2}$. The system remains in this frequency synchronized state until the instability triggers oscillations at $t \approx 10$. At this point the system leaves the unstable frequency-synchronized solution and the phases change at different rates, thus indicating that frequency synchronization is lost until another attractor is reached. The corresponding bottom panel (c) illustrates the temporal evolution of the system, for the same initial set of phases, when control is implemented and turned on. In this case the system does not leave the initial state for sufficiently large time, even though the initial set of phases does not fulfill the condition of the diagonal dominance of the matrix $\mathbf{M}$. The system turns out to be stabilized by the action of control. The same behavior is observed for an initial set of uniformly distributed  phases, see panel (b). After the application of control, we obtain a stable solution which is reminiscent of a traveling wave. Note that for both cases (b) and (d) the corresponding space-time plots with respect to the frequency would look exactly as in Fig 3(b).

\begin{figure}[]
\begin{minipage}[h]{0.99\linewidth}
\center{\includegraphics[width=\linewidth]{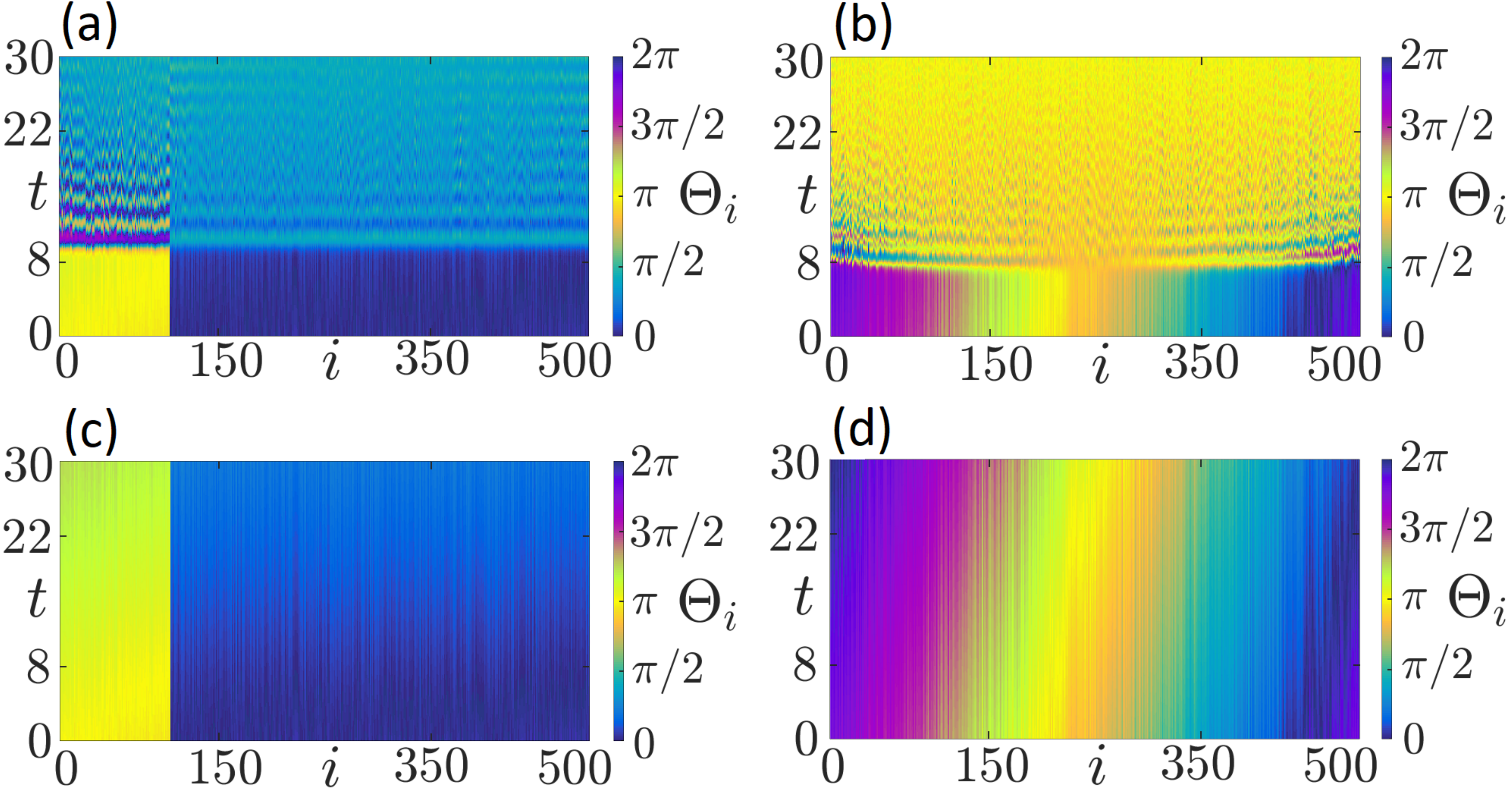}}
\end{minipage}
\caption{Spatio-temporal evolution of phases $\theta_{i}$ without (top panel) and with control (bottom panel). 
Left column: $K=50$ and initial phases $\theta^\ast_{1} =...= \theta^\ast_{150}=\pi$, $\theta^\ast_{151}=...= \theta^\ast_{500} = 0$, (a) control off, $\lambda_{max} = 2.802$; (c) control on, $\lambda_{ctrl} = -0.759$. Right column: $K=70$ and uniformly distributed initial phases, (b) control off, $\lambda_{max} = 2.41$; (d) control on, $\lambda_{ctrl} = -0.823 $. 
Other parameters: as on Fig. 1.}
\label{FIG:4}
\end{figure}

\section{Interplay of noise and disorder}

Finally, we investigate the influence of temporal power fluctuations of generators, which is a signature of renewable energy-based power grids, e.g., wind turbines and photovoltaics. For this purpose we add Gaussian white noise to the network with random connectivity considered so far:

\begin{equation}\label{EQ:28}
m \ddot\theta_i + \dot\theta_i = \Omega_i + \frac{K}{N_c} \sum\limits_{j=1}^N A_{ij} \sin\left(\theta_j - \theta_i\right) + \sqrt{2D}\xi_i (t),
\end{equation}
where $\xi_i$ denotes Gaussian white noise defined by $\langle \xi_i \rangle = 0$ and $\langle \xi_i (t) \xi_j (s) \rangle = \delta_{ij} \delta (t-s)$; $D$ is the noise intensity.

Networks of Kuramoto oscillators with inertia subject to white noise have been recently investigated in \cite{GAM17,SCH17,TUM18} to mimic stochastic power fluctuations typical
for renewable energies, and to compute the synchronization transition scenarios. In particular Tumash et al \cite{TUM18} have investigated the influence of noise on the synchronization transition for globally coupled networks; here we extend these previous studies to random networks. In order to find out how the external noise changes the properties of a diluted network, we investigate the synchronization transition for a random network with average connectivity $N_c= 0.1 \ N$ (i.e., dilution parameter $p=0.10$ and $90 \%$ of links removed)
under the impact of white Gaussian noise of intensity $\sqrt{2D} = 5$, see Fig. 5(a). As expected,  the hysteretic region, identified by the vertical dotted green lines at $K_{c1} = 25$ and $K_{c2} = 5$, becomes 
smaller in comparison to what we observe in Fig.1(a). Moreover the system reveals a smaller K-interval where traveling waves occur in the upsweep. This is consistent with the results found in \cite{TUM18}: 
intermediate values of noise reduce the hysteresis and traveling waves disappear. As the noise is increased, standing waves begin to disappear as well
and intermediate states are not detected any more. Therefore in Fig. 5(a) we observe the typical effect of noise at intermediate intensity while the dilution seems not to play an essential role. 
Panels (b), (c) and (d) of Fig. 5 illustrate the features of the frequency-synchronized state with $\bar{r}=0.96$, corresponding to $K=30$. Panel (b) shows that almost all frequencies are synchronized, while phases and instantaneous phase velocities are not strictly correlated due to the effect of noise, see panels (c) and (d). Moreover both noise and dilution (spatial disorder) contribute to decreasing the value of coupling strength $K$ at which partial frequency synchronization is reached ($K^c_1=25$), as compared to the all-to-all coupled case without noise ($K_{PS}=31$). 

\begin{figure}[]
\begin{minipage}[h]{0.99\linewidth}
\center{\includegraphics[width=\linewidth]{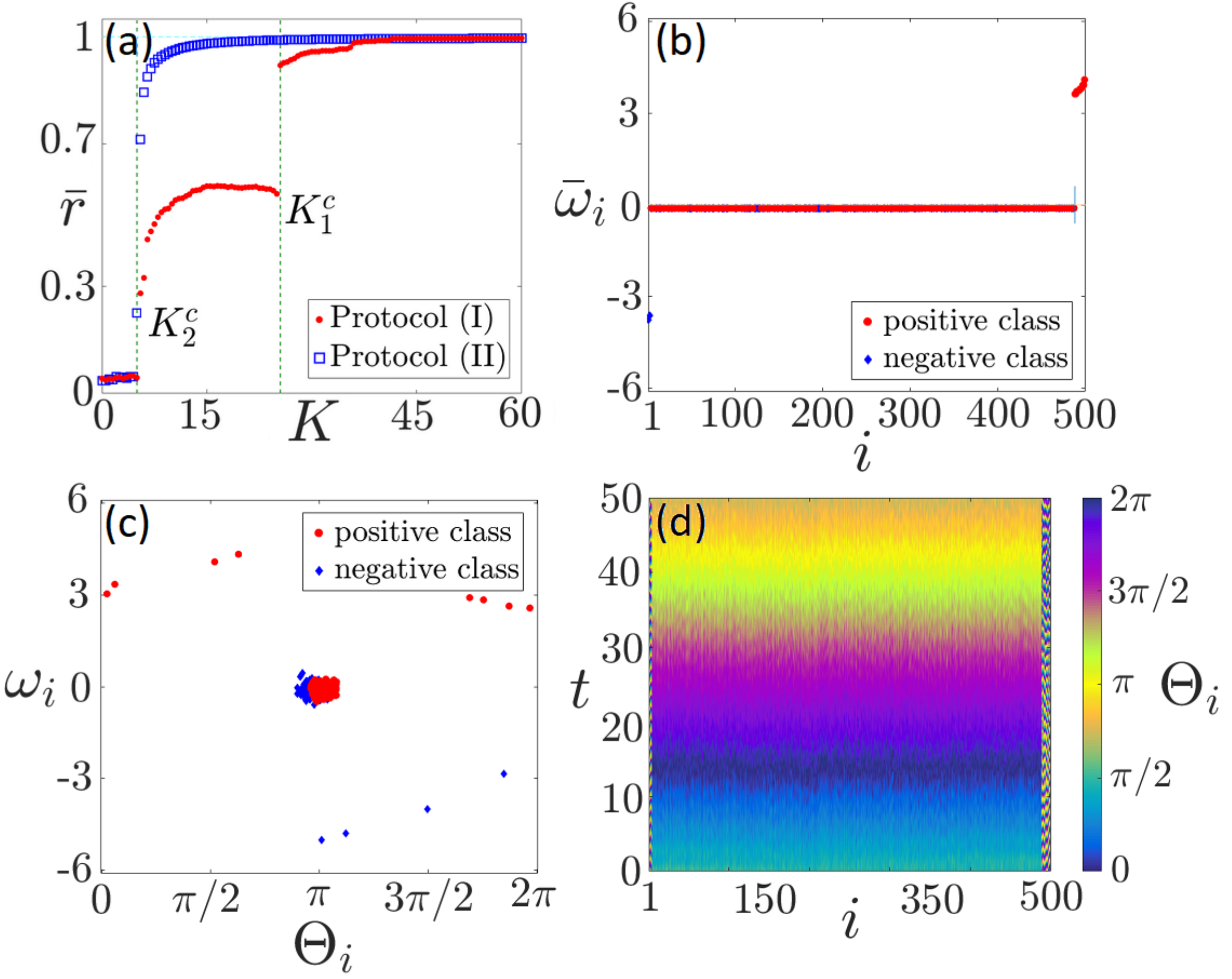}}
\end{minipage}
\caption{(a) Time-averaged global order parameter $\bar{r}$ as a function of coupling constant $K$ for a randomly coupled network with dilution parameter $p=0.10$ and stochastic 
dynamics with noise intensity $\sqrt{2D} = 5$, obtained by following protocol (I) (upsweep, red filled circles) and protocol (II) (downsweep, blue empty squares). The vertical dotted green lines 
denote the boundaries of the hysteretic region. (b) Average phase velocity $\bar{\omega}_i$ for $K=30$ (frequency-synchronized solution). The elements related to 
the positive (negative) distribution of natural frequencies are denoted by red circles (blue diamonds). (c) Instantaneous phase velocity $\omega_i$ versus phase $\theta_i$ for all $i \in \left ( 1 \, ... \, 500 \right )$ (snapshot). (d) Spatio-temporal evolution of phases $\theta_{i}$. Other parameters as in Fig. 1. }
\label{FIG:5}
\end{figure}

\section{Conclusions}

Power grids are typically characterized by sparse networks where nodes have low average connectivity. For this reason we have considered random networks with low average connectivity to model 
the network topology underlying high voltage transmission grids, while the single node dynamics is described in terms of Kuramoto oscillators with inertia. To gain insight into random networks, 
we have studied the synchronization transition for a sparse network, by calculating the time-averaged global order parameter for upsweep and downsweep of the coupling strength. Moreover we have defined 
and characterized the states arising for different levels of synchronization, which mainly differ in the shape of their average phase velocity profile and their average global order parameter: 
asynchronous state, traveling wave, standing wave, and partial (almost complete) synchronization. In particular we have focused on the frequency-synchronized state, since it is mandatory for the 
stable operation of a power grid, irrespectively of its topological connections.

Within this study we have provided mathematical tools which allow us to determine the stability of power grids with random connectivity. We have discussed the lack of phase synchronization in 
networks of heterogeneous rotators, and have derived an expression for the collective frequency at which the system synchronizes when frequency synchronization occurs. In terms of power grids 
we obtain a synchronous solution, required for the stable operations, when the produced power is equal to the consumed power. We have performed a linear stability analysis of the system, 
considering small perturbations around the frequency-synchronized solution and have derived stability criteria, based on the properties of the initial phase differences of the oscillators. 
We have numerically solved the eigenvalue problem using the iterative Levenberg-Marquardt algorithm based on a nonlinear least squares scheme. However, it is not possible to reach a stable 
frequency-synchronized solution for arbitrarily small couplings, therefore we have estimated the critical coupling strength $K_c$ beyond which a frequency-synchronized solution is possible. 
It turns out that the number of possible frequency-synchronized solutions increases with increasing coupling strength. For sufficiently large coupling we can also find unstable solutions that 
are usually characterized by large differences of initial phases. We have provided examples on both unstable and stable solutions for various initial conditions. 
Furtheron, we have implemented a linear feedback control scheme for stabilizing unstable frequency-synchronized solutions for arbitrary initial phases, and all $K>K_c$. Finally, we have briefly 
investigated diluted networks with stochastic dynamics due to temporally fluctuating power. Here we have observed that intermediate noise intensities might play a constructive role in minimizing 
the critical value of coupling strength required to reach partial frequency synchronization. We have also shown that the disorder induced by random connectivities does not drastically change 
the effect of noise, but slightly enhances it.

Future perspectives of this work might be to focus on the initial conditions used as initial guess for the Levenberg-Marquardt algorithm, such that this method becomes more efficient. 
It would also be interesting to investigate whether a stable solution is obtainable for initial phases sufficiently distinct from zero and to investigate in more detail the interplay of noise and 
topology in shaping the functioning of modern power grids.

\begin{acknowledgments}
We acknowledge support from the Deutsche Forschungsgemeinschaft (DFG) in the framework of the SFB 910, Projektnummer 163436311
\end{acknowledgments}


\begin{thebibliography}{10}
\bibitem{TAN97}
H.~A. Tanaka, A.~J. Lichtenberg and S.~Oishi: {\em First order phase transition
resulting from finite inertia in coupled oscillator systems}, Phys.~Rev.~Lett. {\bf 78} (11), 2104 (1997).

\bibitem{TAN97a}
H.~A. Tanaka, A.~J. Lichtenberg and S.~Oishi: {\em  Self-synchronization of coupled
oscillators with hysteretic responses}, Phys.~D: Nonlin.~Phen. {\bf 100} (3), 279-300 (1997). 

\bibitem{ERM91} B. Ermentrout: {\em An adaptive model for synchrony in the firefly Pteroptyx malaccae}, J. Math. Biol.\textbf{29}, 571 (1991).

\bibitem{TRE05} B. R. Trees, V. Saranathan, and D. Stroud: {\em Synchronization in disordered Josephson junction arrays: Small-world connections and the Kuramoto model}, Phys. Rev. E \textbf{71}, 016215 (2005).

\bibitem{JI13}  P. Ji, T. K. DM. Peron, P. J. Menck, F. A. Rodrigues, and J. Kurths: {\em Cluster explosive synchronization in complex networks}, Phys. Rev. Lett. \textbf{110}, 218701 (2013).
 
 \bibitem{FIL08a}
G.~Filatrella, A.~H. Nielsen and N.~F. Pedersen: {\em Analysis of a power grid using
a Kuramoto-like model}, Eur. Phys.~J.~B {\bf 61} (4), 485 (2008) {\bf 5}, 380 (2002).

\bibitem{FRA12}
M.~Frasca, L.~Fortuna and A. Sarra Fiore: {\em A network of oscillators emulating the Italian
high-voltage power grid}, Int. J. Modern Phys., {\bf 26}, 25 (2012).

\bibitem{GAM17}
L.~V. Gambuzza, A.~Buscarino, L.~Fortuna, M.~Porfiri and M.~Frasca: {\em  Analysis of dynamical robustness to noise in power grids}, 
IEEE Journal on Emerging and Selected Topics in Circuits and Systems, {\bf 7}, 3 (2017).

\bibitem{ROH12}
M.~Rohden, A.~Sorge, M.~Timme and D.~Witthaut: {\em Self-organized synchronization
in decentralized power grids}, Phys.~Rev.~Lett. {\bf 109} (6), 064101 (2012).
  
\bibitem{ROH14}
M.~Rohden, A.~Sorge, D.~Witthaut and M.~Timme: {\em Impact of network topology on synchrony of oscillatory power grids}, Chaos {\bf 24}, 013123 (2014).

\bibitem{OLM14a}
S.~Olmi, A.~Navas, S.~Boccaletti and A.~Torcini: {\em Hysteretic transitions in the
Kuramoto model with inertia}, Phys.~Rev.~E {\bf 90} (4), 042905 (2014).

\bibitem{NIS15}
T.~Nishikawa and A.~E. Motter: {\em Comparative analysis of existing models for power- grid synchronization}, New J.~Phys. {\bf 17}, 015012 (2015).

\bibitem{OLM16}
S.~Olmi and A.~Torcini: {\em Dynamics of Fully Coupled Rotators with Unimodal and Bimodal Frequency Distribution}, In: Sch{\"o}ll E., Klapp S., H{\"o}vel P. (eds): 
Control of Self-Organizing Nonlinear Systems. Springer, Berlin  (2016).

\bibitem{ROH17}
M.~Rohden, D.~Witthaut. M.~Timme and H.~Meyer-Ortmanns: {\em Curing critical links in oscillator networks as power flow models}, New Jour.~Phys. {\bf 19}, 013002 (2017).
 
\bibitem{MAR08}
E.~Marris: {\em Energy: Upgrading the grid}, Nature {\bf 454}, 570 (2008). 

\bibitem{TUR99}
J.~A. Turner: {\em A Realizable Renewable Energy Future}, Science {\bf 285}, 687 (1999). 

\bibitem{ACK01}
T.~Ackermann, G.~Andersson and L.~S\"oder: {\em  Distributed generation: a definition}, Elec.~Pow.~Sys.~ Res. {\bf 57}, 195 (2001). 

\bibitem{MIL13}
P.~Milan, M.~W\"achter and J.~Peinke: {\em  Turbulent Character of Wind Energy}, Phys.~Rev.~Lett. {\bf 110}, 13 (2013). 

\bibitem{HEI10}
D.~Heide, L.~von Bremen, M.~Greiner, C.~Hoffmann, M.~Speckmann and S.~Bofinger: {\em  Seasonal optimal mix of wind and solar power in a future, highly renewable Europe}, 
Renewable Energy {\bf 35}, 2483 (2010). 

\bibitem{HEI11}
D.~Heide, M.~Greiner, L.~von Bremen and C.~Hoffmann: {\em  Reduced storage and balancing needs in a fully renewable European power system with excess wind and solar power generation}, 
Renewable Energy {\bf 36}, 2515 (2011). 

\bibitem{ANV16}
M.~Anvari, G.~Lohmann, M.~W\"achter, P.~Milan, E.~Lorenz, D.~Heinemann, M.~Reza Rahimi Tabar and J.~Peinke: {\em Short term fluctuations of wind and solar power systems}, 
New~J.~Phys. {\bf 18} (6), 063027 (2016). 

\bibitem{ANV17}
M.~Anvari, M.~W\"achter and J.~Peinke: {\em Phase locking of wind turbines leads to intermittent power production}, Europhys.~Lett. {\bf 116}, 6 (2017). 

\bibitem{SCH17i}
K.~Schmietendorf, J.~Peinke, and O.~Kamps: {\em The impact of turbulent
  renewable energy production on power grid stability and quality\/}, Eur.
  Phys. J. B {\bf 90}, 222 (2017).

\bibitem{SCH18c}
B.~Sch\"afer, C.~Beck, K.~Aihara, D.~Witthaut, and M.~Timme: {\em
  Non-{G}aussian power grid frequency fluctuations characterized by
  {L}evy-stable laws and superstatistics\/}, Nature Energy {\bf 3}, 119 (2018).

\bibitem{TUM18}
L.~Tumash, S.~Olmi, and E.~Sch{\"o}ll: {\em {E}ffect of disorder and noise in
  shaping the dynamics of power grids\/}, Europhys. Lett. {\bf 123}, 20001
  (2018).

\bibitem{MEH18}
V.~Mehrmann, R.~Morandin, S.~Olmi, and E.~Sch{\"o}ll: {\em Qualitative
  stability and synchronicity analysis of power network models in
  port-{H}amiltonian form\/}, Chaos {\bf 28}, 101102 (2018).

\bibitem{TAH19}
H.~Taher, S.~Olmi, and E.~Sch{\"o}ll: {\em Enhancing power grid synchronization
  and stability through time delayed feedback control\/}  (2019),
  arXiv:1901.05201v1.
	
\bibitem{ACE00}
J.~A. Acebron, L.~L. Bonilla and R.~Spigler: {\em Synchronization in populations of globally coupled oscillators with inertial effects}, Phys.~Rev.~E, {\bf 62} (3), 3437-3454 (2000). 

\bibitem{MIR05}
R.~E. Mirollo and S.~H. Strogatz: {\em The spectrum of the locked state for the Kuramoto model of coupled oscillators}, Phys.~D: Nonlin.~Phenom.  {\bf 205}, 249-266 (2005). 

\bibitem{LEV44}
K.~Levenberg: {\em A Method for the Solution of Certain Non-Linear Problems in Least Squares}, Quarterly of Appl.~Math. {\bf 2}, 164-168 (1944). 

\bibitem{MAR63}
D.~Marquardt: {\em An Algorithm for Least-Squares Estimation of Nonlinear Parameters}, SIAM Journ.~Appl.~Math. {\bf 11} (2), 431-441 (1963). 


\bibitem{SCH17}
B. Sch\"{a}fer, M. Matthiae, X. Zhang, M. Rohden, M. Timme and D. Witthaut: {\em Escape routes, weak links, and desynchronization in fluctuation-driven networks}  Phys. Rev. E, \textbf{95 (6)}, 060203 (2017).

\end{thebibliography}

\bibliographystyle{apsrev4-1}

\end{document}